\documentclass[conference]{IEEEtran}
\IEEEoverridecommandlockouts
\usepackage{cite}
\usepackage{amsmath,amssymb,amsfonts}
\usepackage{algorithmic}
\usepackage{graphicx}
\usepackage{textcomp}
\usepackage{xcolor}
\def\BibTeX{{\rm B\kern-.05em{\sc i\kern-.025em b}\kern-.08em
    T\kern-.1667em\lower.7ex\hbox{E}\kern-.125emX}}

\usepackage{booktabs}

\usepackage{comment}

\usepackage{svg}

\usepackage{bm}

\usepackage{url}

\usepackage{cite}

\usepackage{xcolor}

\usepackage{stackrel}

\usepackage{todonotes} 

\usepackage{lipsum}
\usepackage{algorithm}
\usepackage{siunitx}

\usepackage{epsfig} 
\usepackage[T1]{fontenc}

\usepackage{mathtools}
\usepackage{csquotes}
\usepackage{xcolor}  %

\usepackage{pgfplots} %
\usepackage{adjustbox} %

\usepackage{bbm}
\usepackage{amsbsy}
\usepackage{amssymb}
\usepackage{amsmath}
\usepackage{exscale} %
\usepackage{icomma} %
\usepackage{mathrsfs} %
\usepackage{array} %
\usepackage{multirow} %
\usepackage{listings} %
\usepackage{relsize}
\usepackage[inline, shortlabels]{enumitem}
\usepackage{dsfont}

\newcommand\blfootnote[1]{%
  \begingroup
  \renewcommand\thefootnote{}\footnote{#1}%
  \addtocounter{footnote}{-1}%
  \endgroup
}

\usetikzlibrary{positioning}
\usetikzlibrary{external}
\tikzsetexternalprefix{tikz_fig/}

\begin{document}

\title{Deep Reinforcement Learning for Uplink Multi-Carrier Non-Orthogonal Multiple Access Resource Allocation Using Buffer State Information}

\author{\IEEEauthorblockN{Eike-Manuel Bansbach, Yigit Kiyak and Laurent Schmalen}
\IEEEauthorblockA{Communications Engineering Lab, Karlsruhe Institute of Technology, 76187 Karlsruhe, Germany \\
(email: \texttt{e.bansbach@kit.edu})}
}

\maketitle

\begin{abstract}

For orthogonal multiple access (OMA) systems, the number of served user equipments (UEs) is limited to the number of available orthogonal resources.
On the other hand, non-orthogonal multiple access (NOMA) schemes allow multiple UEs to use the same orthogonal resource.
This extra degree of freedom introduces new challenges for resource allocation.
Buffer state information (BSI), like the size and age of packets waiting for transmission, can be used to improve scheduling in OMA systems.
In this paper, we investigate the impact of BSI on the performance of a centralized scheduler in an uplink multi-carrier NOMA scenario with UEs having various data rate and latency requirements.
To handle the large combinatorial space of allocating UEs to the resources, we propose a novel scheduler based on actor-critic reinforcement learning incorporating BSI.
Training and evaluation are carried out using Nokia's ``wireless suite''.
We propose various novel techniques to both stabilize and speed up training.
The proposed scheduler outperforms benchmark schedulers. 
\end{abstract}

\blfootnote{This work has received funding from the European Research Council (ERC) under the European Union’s Horizon 2020 research and innovation programme (grant agreement No. 101001899). Parts of this work have been funded by the German Federal Ministry of Education and Research (BMBF) within the project Open6GHub (grant agreement 16KISK010).}

\vspace{-5mm}
\section{Introduction}
While in 5G multiple access (MA) is mainly realized using orthogonal MA (OMA), non-orthogonal MA (NOMA) is considered as a key enabling technology to improve the spectral efficiency of next generation's mobile communication networks~\cite{Yuan_NOMA_6G}.
Since orthogonal frequency-division multiple access (OFDMA) hinders the straightforward sharing of a physical resource block (PRB) by multiple user equipments (UEs)~\cite{Road_6G}, the number of orthogonal PRBs limits the number of served UEs~\cite{NOMA_Up_vs_Down}.
In contrast, NOMA allows multiple UEs to use the same PRB~\cite{Road_6G} by cohabitation of UEs in the power domain at the transmitter side and successive interference cancellation (SIC) at the receiver side~\cite{Hojeli}. 
Distinctness of the superposed messages of different UEs can be achieved by either power control of the UE's transmit power or by combining UEs with sufficiently distinct channel gains.
Due to OFDMA's high robustness against frequency-selective fading~\cite{shamsSurveyResourceAllocation2014}, 
multi-carrier NOMA (MC-NOMA) as combination of both techniques~\cite{Wang_MC_NOMA} is considered as an answer to the challenges of the next generation of communication networks~\cite{Road_6G}.

For an OFDMA system with UEs having alike quality of service (QoS) and time-invariant data rate requirements, the resource allocation of PRBs to the UEs can be described as a convex optimization problem.
If the UEs belong to different QoS classes with varying guaranteed bit rates (GBRs), varying maximum packet delay budgets (PDBs) and time-variant data traffic, the optimization problem is non-convex~\cite{shamsSurveyResourceAllocation2014}.
One way to solve this optimization problem is by modelling it as a Markov decision process (MDP) and use deep reinforcement learning (DRL)~\cite{wangDeepReinforcementLearning2019}.
With wireless networks becoming increasingly heterogeneous, DRL-based protocols tailored to specific applications can outperform general-purpose solutions~\cite{Mota_MARL}. 
We discussed several DRL-based approaches for scheduling in OFDMA systems in~\cite{bansbach21}.
The approach proposed in~\cite{bansbach21} uses buffer state information (BSI), i.e., the size and age of packets inside the buffer waiting for transmission.
It is able to handle up to 32 UEs belonging to different QoS classes and outperforms non-DRL-based benchmark algorithms.
However, NOMA capabilities aren't considered. \\
Since NOMA increases the number of possible combinations when allocating resources, a DRL algorithm that can handle a large action space is necessary. 
While~\cite{Silva_Noma_Q} uses Q-learning for NOMA resource allocation, all devices belong to the same QoS class and have time-invariant data traffic.
In~\cite{Lillicrap}, the deep deterministic policy gradient (DDPG) algorithm, capable of dealing with an infinite number of DRL actions, e.g., allocation options, is introduced.
While~\cite{He_DQN_DDPG} and~\cite{Wang_multi_cell_noma} use Q-learning for PRB assignment and DDPG for power allocation,~\cite{Wang_MC_NOMA} and~\cite{Xu_V2V_DDPG} directly use DDPG algorithm for PRB assignment.
In~\cite{Wang_MC_NOMA}, a flexible amount of users can be allocated to a PRB for a downlink MC-NOMA scenario. 
This DDPG-based approach outperforms various benchmarks.
In~\cite{Xu_V2V_DDPG}, an uplink MC-NOMA system is investigated.
For time-varying data and two QoS classes, the DDPG-based approach outperforms a Q-learning based reference.
In~\cite{Wang_MC_NOMA} BSI is neglected, in~\cite{Xu_V2V_DDPG} BSI is reduced to the buffer queue length. \\ 
In this work, we extend our work of~\cite{bansbach21} towards a DDPG algorithm for uplink MC-NOMA resource allocation with full BSI.
All UEs have alike transmit power and we ensure discriminability by learning to combine UEs with sufficiently distinct channel gains.
Our main contributions are the inclusion of BSI into the uplink MC-NOMA resource allocation as well as, to the best of our knowledge, the first DRL-based algorithm for uplink MC-NOMA resource allocation without power control of the UEs' transmit powers.

\section{Uplink MC-NOMA System Model}
The used system model is the \textit{NomaULTimeFreq-ResourceAllocation-v0} environment provided by Nokia's ``Wireless Suite'' problem collection~\cite{NokiaWirelesssuite2021}.
It simulates an uplink MC-NOMA scenario and provides benchmarks. 
The set of $K$ UEs is given by $\mathcal{K}=\{\text{UE}_1,\ldots \text{UE}_K\}$ and the set of~$N$ PRBs by $\mathcal{N}=\{\text{PRB}_1,\ldots \text{PRB}_N\}$.
Each UE has a buffer with $L$ slots, which contain packets waiting for transmission.
The UEs belong to one of four QoS classes, identified by their corresponding QoS identifier (QI) $\mathfrak{q}_k \in \{1,2,3,4\}$. 
The QoS classes are given as GBR services, like conversational voice, conversational video and delay critical GBR, and as non-GBR services, like web browsing. 
Upon initialization of an environment, UEs are randomly spread over a $\SI{1}{\kilo\metre}^2$ squared area and assigned a QI.
The area is an empty Euclidean space with a transceiver base station (BS) at its center.
The UEs roam around at rectilinear trajectories with random speeds~\cite{valcarceTimeFreqResourceAllocationv0Environment2020}.
The speeds are independently sampled from a normal distribution according to~\cite{chandraSpeedDistributionCurves2013}.
At the egdes of the area, the UEs bounce off at specular angles~\cite{valcarceTimeFreqResourceAllocationv0Environment2020}. \\
The environment is described by Alg. \ref{alg:env}.
First, the UEs are placed randomly across the area (initialization).
At the beginning of every environment step, the scheduler receives the channel quality indicator (CQI) $\mathfrak{c}_k \in \{0,\ldots 15\}$ of every UE as well as the ages and sizes of the packets inside a UE's buffer.
The scheduler assigns up to $M$ UEs to each PRB, where $M$ is the number of NOMA capabilities of a single PRB. 
Thus, up to $M$ UEs can do NOMA using a single PRB.
The possible actions for each NOMA resource of a PRB are either assigning one of the $K$ UEs or leave it empty, resulting in $K+1$ possible actions.
For every chosen UE, the achievable rate is calculated by 
\begin{align*} 
    R = \frac{B}{N} \log_2(1+\text{SINR}) \; ,
\end{align*}
with the overall system bandwidth $B$.
The UE's signal-to-interference-plus-noise-ratio SINR is calculated by 
\begin{align*} 
    \text{SINR} = \frac{P_{\text{rx}}}{\sigma^2_\text{n}+\zeta+\xi} \; ,
\end{align*} 
where $P_{\text{rx}}$ is the UE's power received at the BS, $\sigma^2_\text{n}$ the power of additive white Gaussian noise, $\zeta\hat{=} -105\,\text{dBm}$ a constant interference power throughout the coverage area and~$\xi$ the interference caused by other UEs occupying the current PRB. 
$P_{\text{rx}}$ is calculated using the distance~$d$ between the UE and the BS, applying the free-space loss to the transmit power~$P_\text{tx} = 13\;\text{dBm}$, which is identical for all UEs.
After each PRB is assigned and the transmitted bits are deleted from the buffers, it is checked whether the remaining packets exceed their latency requirements, given by their PDB.
The PDB depends on the UE's QoS class.
If there are packets that exceed their PDB, the environment returns a negative reward (penalty) by summing up the bits of packets exceeding their PDB~\cite{valcarceTimeFreqResourceAllocationv0Environment2020}.
Afterwards, the UEs move and new packets are generated.  
While SIC demodulates the signals of UEs in the order of decreasing received power and eliminates the interfering waveforms one at a time~\cite[Sec. 16.3-4]{Proakis2007}, the \textit{NomaULTimeFreq-ResourceAllocation-v0} environment emulates SIC by stepwise adding interference.
The reward function is predefined by the environment~\cite{valcarceTimeFreqResourceAllocationv0Environment2020} and is not modified within this paper.

\begin{algorithm}[t]
    \caption{Pseudocode of \textit{NomaULTimeFreqResourceAllo-cation-v0} environment~\cite{NokiaWirelesssuite2021}}
    \label{alg:env}
    \begin{algorithmic}[]
        \STATE Initialize: $K$ UEs with random QI, $N$ PRBs, $M$ UEs per PRB, $L$ buffer slots per UE, maximum runtime $T$ 
        \STATE Randomly place UEs, initialize speed, generate packets
        \FOR{$t = 1,\ldots T_\text{max}$}
            \STATE Get CQI and BSI of all UEs
            \FOR{$n=1,\ldots N$}
                \STATE Choose up to $M$ UEs to do NOMA on $\text{PRB}_n$
                \STATE Set of chosen UEs: $\mathcal{D}\subseteq\mathcal{K},\,|\mathcal{D}|\leq M$
                \STATE NOMA Interference: $\xi=0$
                \FOR{$\text{UE} \in \mathcal{D}$}   
                    \STATE Calculate receive power $P_\text{rx}$ and achievable rate $R$
                    \STATE Transmit packets, delete from UE's buffer
                    \STATE $\xi\leftarrow \xi+P_{\text{rx}}$
                \ENDFOR 
                \STATE Update BSI
            \ENDFOR 
            \STATE Calculate penalty by checking packet ages~\cite{valcarceTimeFreqResourceAllocationv0Environment2020} 
            \STATE Move UEs, generate new packets
        \ENDFOR
    \end{algorithmic}
\end{algorithm}

While the OFDMA resource allocation problem in~\cite{bansbach21} focuses on scheduling by the packet's urgency, the MC-NOMA problem extends the OFDMA problem by combining different UEs on the same PRB, while mitigating the interference among the UEs.  

\section{Reinforcement Learning}
\subsection{Reinforcement Learning Problem}
In RL, an agent tries to skillfully map actions to observed system states in order to maximize a numerical reward~\cite{bansbach21}.
Starting at an observed state, the agent takes an action, receives a reward and follow-up state, takes another action, et cetera. 
This sequential decision making can be formalized as an MDP~\cite[Chap.~3]{suttonReinforcementLearningIntroduction2018}.
Let $\mathcal{S}$ denote a finite set of states, $\mathcal{A}$ a finite set of actions and $S_t \in \mathcal{S}$, $A_t\in \mathcal{A}$ as well as $W_t \in \mathbb{R}$ random variables describing the state, action and reward at time~$t$.
Assuming the Markov property is fulfilled, the dynamics of an MDP can be fully described by the state-transition probabilities $p(S_{t+1}=s'|S_t=s,\,A_t=a): \mathcal{S}\times \mathcal{S} \times \mathcal{A} \rightarrow [0,1]$ with $s',s \in \mathcal{S}$ and $a\in\mathcal{A}$, as well as the expected reward $w(S_t=s,\,A_t=a,\,S_{t+1}=s'):\mathcal{S} \times \mathcal{A} \times \mathcal{S} \rightarrow \mathbb{R}$ when taking action $A_t=a$ at state $S_t=s$ with follow-up state~$S_{t+1}=s'$~\cite[Sec. 3.1]{suttonReinforcementLearningIntroduction2018}. \\
With $ G_t := \sum_{k=0}^\infty \gamma^k W_{t+k+1}\, , \gamma \in [0,1]$ as the discounted return,
the cumulative reward in the long run is calculated~\cite[Sec.~3.3]{suttonReinforcementLearningIntroduction2018}. 
A policy $\pi(a|s)$ defines probabilities of taking action $A_t=a$ given state $S_t=s$ in order to maximize $G_t$~\cite[Sec~3.4]{suttonReinforcementLearningIntroduction2018}.
Following a policy $\pi$, the action-value function $q_\pi(s,a)$ denotes the expected return for choosing an action $A_t=a$ at state $S_t=s$~\cite[Sec.~3.5]{suttonReinforcementLearningIntroduction2018}:
\begin{align*}
    q_\pi(s,a) := \mathbb{E}_\pi\left\{G_t|S_t=s,A_t=a \right\} \quad \forall s \in \mathcal{S},\,\forall a \in \mathcal{A}\, .
\end{align*}
A policy $\pi$ is called an optimal policy $\pi^*$, if its decisions maximize the action-value function 
\begin{align*} 
    q^*(s,a) := \max_{\pi} q_\pi(s,a) \quad \forall s \in \mathcal{S},\,\forall a \in \mathcal{A} \, ,
\end{align*}
where $q^*(s,a)$ is called the optimal action-value function~\cite[Sec.~3.6]{suttonReinforcementLearningIntroduction2018}. 
Similarly, the value function $v_\pi(s)$ as the expected return when in $S_t=s$ and following $\pi$ can be defined by
$v_\pi(s) := \mathbb{E}_\pi \left\{ G_t|S_t=s \right\} \, \forall s \in \mathcal{S}$ ~\cite[Sec.~3.5]{suttonReinforcementLearningIntroduction2018}.
The RL problem can be solved using value-based or policy-based methods.
While the former try to learn $q^*(s,a)$, the latter directly optimize the policy $\pi(a|s)$~\cite[Chap.~13]{suttonReinforcementLearningIntroduction2018}. 

\subsection{Value-based Methods}
The objective of Q-learning is to use temporal difference learning to learn a function $Q(s,a)$ that approximates $q^*(s,a)$~\cite{watkinsQlearning1992}.
Performing action $A_t=a$ at state $S_t=s$ and observing the reward $W_{t+1}=w$ and follow-up state $S_{t+1}=s'$ yields the tuple $(s,a,w,s')$.
The approximated Q-function $Q(s,a)$ can be updated using 
\begin{align*} 
    Q(s,a) \leftarrow (1-\alpha)Q(s,a) + \alpha[w+\gamma \max_{a'}Q(s',a')] \; ,
\end{align*}
with the step-size $\alpha \in [0,1]$~\cite{watkinsQlearning1992}.
Using a more accurate estimate incorporating the observed reward $w$ for state-action pair $(s,a)$, $Q(s,a)$ gets updated.
Q-learning converges to the optimal action-value function  $q^*(s,a)$~\cite{watkinsQlearning1992}.
The optimal action $A_t=a^*$ given state $S_t=s$ is chosen greedily by $a^* = \underset{a \in \mathcal{A}}{\text{arg}\,\text{max}}\, Q(s,a)$ ~\cite{watkinsQlearning1992}.

\subsection{Policy-based Methods}
Instead of learning $q^*(s,a)$, policy-based methods directly optimize the policy parameters $\bm{\theta} \in \mathbb{R}^d$ of a parametrized policy $\pi_{\bm{\theta}}:= \pi(a|s,\bm{\theta})$.
With the performance measure ${J(\bm{\theta}):=v_{\pi_{\bm{\theta}}}(s_0)}$ as the expected return when starting at state $S_0 =s_0$ and following policy $\pi_{\bm{\theta}}$, the policy can be updated using \emph{gradient ascent}~\cite[Chap.~13]{suttonReinforcementLearningIntroduction2018}.
Following the stochastic policy gradient theorem (PGT), it can be shown that~\cite{Policy_theorem}
\begin{align*}
    \nabla J(\bm{\theta}) %
    &\propto \mathbb{E}_\pi \left[ \sum_a  q_\pi(S_t,a) \; \nabla_{\bm{\theta}} \pi(a|S_t,\bm{\theta}) \right] \; .
\end{align*}
$\nabla J(\bm{\theta})$ is proportional to the expectation of the gradient of the probability choosing action $a$ given state $S_t$, weighted with its expected return $q_\pi(s,a)$.
Hence, the computation of the performance gradient is reduced to a simple expectation~\cite{silver_DDPG}. \\
Using stochastic gradient ascent, $\bm{\theta}$ can be updated by
\begin{align*} 
        \bm{\theta} \leftarrow \bm{\theta} + \epsilon \sum_a \hat{q}(S_t,a)\; \nabla_{\bm{\theta}} \pi(a|S_t,\bm{\theta}) \; ,
\end{align*}
where $\epsilon$ is a step-size and $\hat{q}(S_t,a)$ is a learned approximation of $q_\pi(s,a)$~\cite[Sec.~13.2]{suttonReinforcementLearningIntroduction2018}. 
Hence, policy-gradient methods require an estimate of the action-value function $q_\pi(S_t,a)$~\cite{silver_DDPG}.

\subsection{Actor-critic Methods}
\begin{figure}
    \centering
    \begin{tikzpicture}[>=latex]
    \node[] (base) {};
    \node[draw, thick, right = 2.1cm of base, rectangle, rounded corners, minimum height=0.7cm, minimum width=2.5cm] (env) {Environment};
    \node[draw, thick, above=0.2cm of base,rectangle, rounded corners,  minimum height=0.7cm, minimum width=2.5cm] (critic) {Critic};
    \node[draw, thick, below=0.1cm of base,rectangle, rounded corners,  minimum height=0.7cm, minimum width=2.5cm] (actor) {Actor};
    
    \node[draw, circle, fill=black, left = 0.3cm of env, inner sep=1pt] (dot1) {}; 
    \node[draw, circle, fill=black, left = 0.5cm of actor, inner sep=1pt] (dot2) {}; 
    \node[draw, thick, rounded corners, rectangle,below= 0.9cm of env] (z) {$z^{-1}$};
    \node[left=0.5pt of dot1] (a_t) {$A_t$};

    \draw[->,thick] (env)++(1.24cm,-0.1cm) -- ++(0.5cm,0cm) |- (z) node[pos=0.75,above] {$S_{t+1}$};
    \draw[thick] (z) -| (dot2) node[pos=0.05,above] {$S_{t}$};
    \draw[->,thick] (dot2) -- (actor);
    \draw[->,thick] (dot2) |- (critic);

    \draw[thick] (actor) -| (dot1);
    \draw[->,thick] (dot1) -- (env);
    \draw[->,thick] (dot1) |- ++(-0.62cm,0.55cm);

    \draw[->,thick] (env)++(1.24cm,0.1cm) -| ++(0.5cm,0.72cm) -- ++(-3.98cm,0cm) node[pos=0.39,above] {$W_{t+1}$};

    \draw[dotted,->,thick] (critic)++(0cm,-0.36cm) -- ++(0cm,-0.3cm) -- ++(-0.5cm,0cm) -- ++(0cm,0.1cm) -- ++(1.2cm,1.2cm);
    \draw[dotted,->,thick] (critic)++(0cm,-0.36cm) -- ++(0cm,-1.65cm);

    \node[draw, fill=white, thick, above=0.2cm of base,rectangle, rounded corners,  minimum height=0.7cm, minimum width=2.5cm] (Critic) {critic  $Q(s,a|\bm{w})$};
    \node[draw, fill=white, thick, below=0.1cm of base,rectangle, rounded corners,  minimum height=0.7cm, minimum width=2.5cm] (Actor) {actor $\pi_{\bm{\theta}}(s)$};

\end{tikzpicture}
    \caption{Schematic overview of an actor-critic algorithm adapted from~\cite[Fig.~1]{actor-critic_survey}. The dotted lines indicate that the critic is responsible for updating the actor and itself. }
    \label{fig:actor_critic_env}
\end{figure}
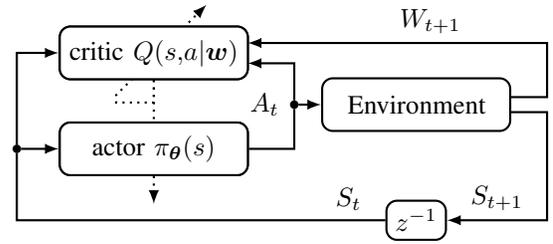

While policy-based methods are able to handle large or continuous action spaces, value-based methods have a lower variance in the estimates of expected returns.
Actor-critic methods combine the advantages of both~\cite{actor-critic_survey}.
A parametrized \emph{actor}~$\pi_{\bm{\theta}}(s)$ with parameters $\bm{\theta}$ is defined.
Using, e.g., Q-learning, a \emph{critic} with parameters $\bm{\Phi}$ estimates the action-value function $Q(s,a|{\bm{\Phi}})\approx q_{\pi}(s,a)$ of the actor's policy $\pi_{\bm{\theta}}(s)$.
As depicted in Fig. \ref{fig:actor_critic_env}, both the actor and the critic receive a state~$S_t$.
The actor chooses an action $A_t$ following~$\pi_{\bm{\theta}}(s)$.
Using the reward~$W_{t+1}$, the critic updates its estimator $Q(s,a|{\bm{\Phi}})$.
Afterwards, $Q(s,a|{\bm{\Phi}})$ is used to update the actor according to the stochastic PGT update step~\cite{Konda_actor_critic}.

\subsection{Deterministic Policy Gradient}
Now consider a deterministic parametrized policy $\rho_{\bm{\theta}}(s)$ with parameters $\bm{\theta}$, which maps exactly one action to every state.
Given its performance measure $J(\bm{\theta}):=v_{\rho_{\bm{\theta}}}(s_0)$ when starting at state $S_0=s_0$, the deterministic PGT is given by~\cite{silver_DDPG}
\begin{align*}
    \nabla J(\bm{\theta}) = \mathbb{E}_{\rho} \left[ \nabla_{\bm{\theta}} \rho_{\bm{\theta}}(S_t) \; \nabla_a q_{\rho}(S_t,a) |_{a=\rho_{\bm{\theta}}(S_t)} \right] \, ,
\end{align*}
where $q_{\rho}(s,a)$ is the action-value function following policy~$\rho_{\bm{\theta}}$.
Compared to the stochastic PGT, deterministic PGT updates only for the taken action $a=\rho_{\bm{\theta}}(S_t)$.  
Moreover, instead of $q_{\rho}(s,a)$, only its gradient $\nabla q_{\rho}(s,a)$ is taken into account, which avoids the estimation of $q_{\rho}(s,a)$ and is computationally attractive. 
The stochastic policy gradient converges to the deterministic policy gradient for $\sigma^2 \rightarrow 0$, where $\sigma^2$ is the output variance of the stochastic policy ~\cite{silver_DDPG}. \newline

\subsection{Deep Deterministic Policy Gradient Algorithm}
In~\cite{Lillicrap}, the DDPG algorithm, an actor-critic setup using deterministic PGT, is introduced.
Since feedforward neural networks define a parametrized mapping $\bm{y}=f(\bm{x},\bm{\theta})$, RL algorithms can be implemented using deep neural networks (DNNs)~\cite[Sec.~9.7]{suttonReinforcementLearningIntroduction2018}.
However, there are two challenges when using DNNs~\cite{Lillicrap}: 
First, most DNN optimization algorithms assume that the samples used for optimization are i.i.d..
Storing observed transitions $(s_t,a_t,s_{t+1},r_{t+1})$ in a finite replay memory and randomly sampling from it ensures learning using (approximately) independent transitions.
Second, using only a single DNN for, e.g., Q-learning, updating the DNN while using it for calculating the target $Q(s',a')$ destabilizes learning.
Using a copy of the networks, the actual network is updated and the copy, called target, is used to estimate $Q(s',a')$.
The target network gets periodically synchronized with the updated network.  
The full algorithm is given by~\cite[Alg.~1]{Lillicrap}.

\section{Actor-Critic Methods for NOMA-OFDMA Uplink Resource Allocation}
\subsection{Sequential NOMA Allocation}
\begin{figure}
    \centering
    \resizebox{0.48\textwidth}{!}{
    \begin{tikzpicture}[>=latex]
    
    \node[draw,rectangle,rounded corners,minimum width=1.5cm] (ac1) {\large Actor};
    \node[draw,rectangle,rounded corners,right=1.2cm of ac1,minimum width=1.5cm] (ac2) {\large Actor};
    \node[right=0.2cm of ac2] (dots) {\LARGE$\cdots$};
    \node[draw,rectangle,rounded corners,right=1cm of dots,minimum width=1.5cm] (acM) {\large Actor};

    \node[draw,rectangle,rounded corners,above=1cm of ac1,minimum width=1.5cm] (X1) {$\bm{X}\leftarrow \bm{x}_1$};
    \node[draw,rectangle,rounded corners,above=1cm of ac2,minimum width=1.5cm] (X2) {$\bm{X}\leftarrow \bm{x}_2$};
    \node[draw,rectangle,rounded corners,above=1cm of acM,minimum width=1.5cm] (XM) {$\bm{X}\leftarrow \bm{x}_M$};
    \node[right=0.4cm of XM] (X) {$\bm{X}$};

    \draw[->,thick] (ac1) -- (X1) node[pos=0.5,right] {$\bm{x}_1$};
    \draw[->,thick] (ac2) -- (X2) node[pos=0.5,right] {$\bm{x}_2$};
    \draw[->,thick] (acM) -- (XM) node[pos=0.5,right] {$\bm{x}_M$};

    \node[below=1cm of ac1, circle, inner sep=1pt, draw, fill=black] (p1) {};
    \node[below=1cm of ac2, circle, inner sep=1pt, draw, fill=black] (p2) {};
    \node[left=1.3cm of p1] (S) {$\bm{S}_t$};
    \node[right=1.33cm of p2] (helper1) {};
    \node[right=0.55cm of helper1] (helper2) {};

    \draw[->,thick] (S) -| (ac1);
    \draw[->,thick] (p1) -| (ac2);
    \draw[thick] (p2) -- (helper1);
    \draw[dotted,thick] (p2) -- ++(3.5cm,0cm);
    \draw[->,thick] (helper2) -| (acM);

    \node[left=0.4cm of ac1] (X0) {$\bm{X}=\bm{0}$};
    \draw[->,thick] (X0) -- (ac1);
    \draw[->,thick] (X1) -- ++(1.5cm,0cm) |- (ac2);
    \draw[thick] (X2) -- ++(1.4cm,0cm);
    \draw[thick,dotted] (X2) -- ++(1.4cm,0cm) -- ++(0.8cm,0cm) |- (acM);
    \draw[->,thick] (dots)++(1cm,0cm) -- (acM);
    \draw[->,thick] (XM) -- (X);
\end{tikzpicture}
    }
    \caption{Sketch of the proposed decision process of scheduling the NOMA resources of one PRB.
    The decision matrix is initialized as~${\bm{X}=\bm{0} \in \mathbb{R}^{M \times (K+1)}}$ and is updated row-wise using the action probabilities~${\bm{x}_m \in \mathbb{R}^{(K+1)}}$ of NOMA-resource~$m$.
    }
    \label{fig:actor_decision}
\end{figure}
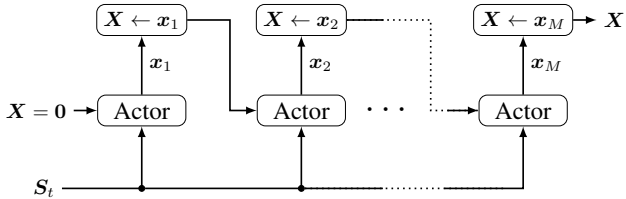

While the general DRL framework introduced in Sec.~III receives~$S_t$, takes~$A_t$ and immediately receives~$W_{t+1}$ and~$S_{t+1}$, the uplink MC-NOMA scenario in Alg. \ref{alg:env} involves two major challenges.
First, the BSI and, therefore, the state~$S_t$ is updated after all NOMA-resources of a PRB are allocated.
Thus, multiple sequential allocation actions, allocating the NOMA-resources of a PRB, need to be taken without new state information.
If, e.g., UE~$k$ is allowed to use the first NOMA-resource and is able to transmit all the packets inside its buffer, allocating the second NOMA resource to UE~$k$ does not make sense.
However, the state containing information about the occupancy of all UEs' buffers is not updated and still assumes that UE~$k$ has packets to transmit.
To combat this issue, we introduce a sequential decision making structure, shown in Fig.~\ref{fig:actor_decision}, which is inspired by~\cite{Zang_actor_setup}.
The rows~${\bm{x}_m\in \mathbb{R}^{(K+1)},\; m=\{1,\ldots M\}}$, of the matrix~${\bm{X} \in \mathbb{R}^{M\times (K+1)}}$ contain the probabilities of the~$K+1$ actions to be chosen for NOMA resource~$m$.
So,~$\bm{X} = (\bm{x}_1 \ldots \bm{x}_M)^\mathrm{T}$ with~$\bm{x}_m=(x_{m,1}\ldots x_{m,K+1})$ and~$x_{m,k} \in [0,1]$.  
We initialize~$\bm{X}=\bm{0}$ and iteratively decide for allocation, replacing the~$m^\text{th}$ row of~$\bm{X}$ with the actor output~$\bm{x}_m$. 
The updated~$\bm{X}$ is used to enable prediction of the changes applied to the state, e.g., which UE's buffers may have been emptied using already allocated NOMA-capabilites.
After the actor decided for~$\bm{x}_M$, the actions~$\bm{a} = (a_1 \ldots a_M)$ are sampled from the probability distributions provided by~$\bm{x}_m$.  

\subsection{Early Termination}
The second issue we face in the uplink MC-NOMA scenario are sparse rewards.
While the state is updated after each PRB is processed, the rewards are calculated after all PRBs are processed.
Thus,~$M\cdot N$ actions have been carried out.
Not getting rewards immediately on single actions, but on a group of actions, prolongs training. 
We combat sparse rewards by introducing early termination for training.
Especially at the beginning of training, the actor decides for suboptimal actions, leading to full UE buffers and, therefore, to situations without a chance for choosing beneficial actions.
We assume that a sufficiently trained agent is able to avoid such ill-conditioned situations.
Therefore, we terminate a training episode early if the training reward~$W_{\text{train}}<W_{\text{cap}}$, where~$W_{\text{cap}} \in \mathbb{R}$ is an empirically determined value.
Since the reward penalizes packets that exceeded their PDB (punishment), the reward is solely negative~$W_{t+1}\leq 0$ (penalty-only) and it is~$W_{\text{train}}<0$.  
With early termination, we save computation time in ill-conditioned situations and are able to combat the prolonged training due to sparse rewards.

\subsection{Traffic-based Masking}
To avoid the actor choosing UEs with empty buffers, we introduce traffic-based masking.
By defining a Boolean mask~$\bm{h} \in \{0,1\}^{(K+1)}$, indicating whether a UE has packets in its buffer or not, the probability vector~$\bm{x}_m$ is updated by an elementwise multiplication~$\bm{x}_m \leftarrow \bm{x}_m \odot \bm{h}$. 
We apply normalization to ensure that after masking~$\bm{x}_m$ is a probability distribution. 

\subsection{Architecture of Actor and Critic Networks}
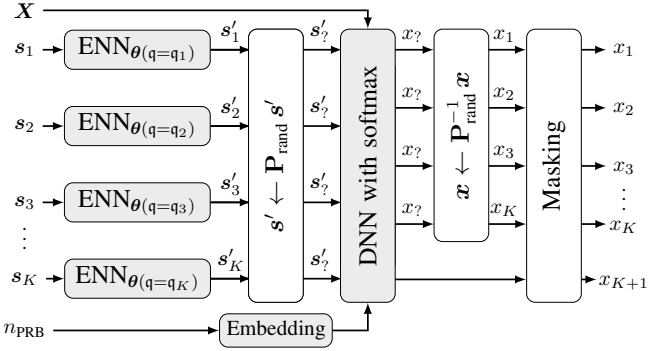
\begin{figure}
    \centering
    \resizebox{0.5\textwidth}{!}{
        \begin{tikzpicture}[>=latex]
    \node[] (x1) {$\bm{s}_1$};
    \node[below=0.8cm of x1] (x2) {$\bm{s}_2$};
    \node[below=0.8cm of x2] (x3) {$\bm{s}_3$};
    \node[below=0.8cm of x3] (x4) {$\bm{s}_K$};
    \path (x3) -- (x4) node [black, midway,sloped] {$\dots$};

    \node[draw, fill=gray!15,right=10pt of x1,rectangle, rounded corners,minimum width=2.4cm] (ENN1) {\mbox{\large$\text{ENN}_{\boldsymbol{\theta}(\mathfrak{q}=\mathfrak{q}_1)}$}};
    \node[draw, fill=gray!15,right=10pt of x2,rectangle, rounded corners,minimum width=2.4cm] (ENN2) {\mbox{\large$\text{ENN}_{\boldsymbol{\theta}(\mathfrak{q}=\mathfrak{q}_2)}$}};
    \node[draw, fill=gray!15,right=10pt of x3,rectangle, rounded corners,minimum width=2.4cm] (ENN3) {\mbox{\large$\text{ENN}_{\boldsymbol{\theta}(\mathfrak{q}=\mathfrak{q}_3)}$}};
    \node[draw, fill=gray!15,right=8.5pt of x4,rectangle, rounded corners,minimum width=2.4cm] (ENN4) {\mbox{\large$\text{ENN}_{\boldsymbol{\theta}(\mathfrak{q}=\mathfrak{q}_K)}$}};

    \draw[->,thick] (x1) -- (ENN1);
    \draw[->,thick] (x2) -- (ENN2);
    \draw[->,thick] (x3) -- (ENN3);
    \draw[->,thick] (x4) -- (ENN4);

    \node[below=0.2cm of ENN2] (middle) {};
    \node[right =1.7cm of middle,draw, rounded corners,minimum height=4.5cm, minimum width=0.9cm,label={[rotate=90]center:\mbox{\large$\bm{x} \leftarrow \bm{P}_\text{rand}\,\bm{x} $}}] (UE shuf) {};

    \node[right=0.6cm of UE shuf, draw, rounded corners, fill=gray!15, minimum height=3.5cm, minimum width=0.9cm, label={[rotate=90]center:\mbox{\large DQN}}] (DQN) {};

    \node[above=5pt of  x1] (actor) {$\bm{X}$};
    \draw[->,thick] (actor) -| ++(5.65cm,-0.3cm);

    \draw[->,thick] (ENN1) -- ++(1.87cm,0) node[pos=0.5,above] {$\bm{s}_1'$};
    \draw[->,thick] (ENN2) -- ++(1.87cm,0) node[pos=0.5,above] {$\bm{s}_2'$};
    \draw[->,thick] (ENN3) -- ++(1.87cm,0) node[pos=0.5,above] {$\bm{s}_3'$};
    \draw[->,thick] (ENN4) -- ++(1.87cm,0) node[pos=0.5,above] {$\bm{s}_K'$};

    \draw[->,thick] (ENN1) -- ++(3.37cm,0) node[pos=0.83,above] {$\bm{s}_?'$};
    \draw[->,thick] (ENN2) -- ++(3.37cm,0) node[pos=0.83,above] {$\bm{s}_?'$};
    \draw[->,thick] (ENN3) -- ++(3.37cm,0) node[pos=0.83,above] {$\bm{s}_?'$};
    \draw[->,thick] (ENN4) -- ++(3.37cm,0) node[pos=0.83,above] {$\bm{s}_?'$};

    \draw[->,thick] (ENN1)++(4cm,0cm) -- ++(0.88cm,0) node[pos=0.6,above] {$x_?$};
    \draw[->,thick] (ENN2)++(4cm,0.3cm) -- ++(0.88cm,0) node[pos=0.6,above] {$x_?$};
    \draw[->,thick] (ENN3)++(4cm,0.6cm) -- ++(0.88cm,0) node[pos=0.6,above] {$x_?$};
    \draw[->,thick] (ENN4)++(4cm,0.9cm) -- ++(0.88cm,0) node[pos=0.6,above] {$x_?$};

    \draw[->,thick] (ENN1)++(5.5cm,0cm) -- ++(0.9cm,0) node[pos=0.6,above] {$x_1$};
    \draw[->,thick] (ENN2)++(5.5cm,0.3cm) -- ++(0.9cm,0) node[pos=0.6,above] {$x_2$};
    \draw[->,thick] (ENN3)++(5.5cm,0.6cm) -- ++(0.9cm,0) node[pos=0.6,above] {$x_3$};
    \draw[->,thick] (ENN4)++(5.5cm,0.9cm) -- ++(0.9cm,0) node[pos=0.6,above] {$x_K$};

    \draw[->,thick] (ENN1)++(7cm,0cm) -- ++(0.72cm,0);
    \draw[->,thick] (ENN2)++(7cm,0.3cm) -- ++(0.72cm,0);
    \draw[->,thick] (ENN3)++(7cm,0.6cm) -- ++(0.72cm,0);
    \draw[->,thick] (ENN4)++(7cm,0.9cm) -- ++(0.72cm,0);

    \node[right=6.5cm of ENN1] (q1) {$x_1$};
    \node[below=0.52cm of q1] (q2) {$x_2$};
    \node[below=0.52cm of q2] (q3) {$x_3$};
    \node[below=0.52cm of q3] (q4) {$x_K$};
    \path (q3) -- (q4) node [black, midway,rotate=90] {$\dots$};
    \node[below=0.45cm of q4] (q5) {$x_{K+1}$};

    \draw[->,thick] (q5)++(-3.81cm,0.05cm) -- ++(2.2cm,0cm);
    \draw[->,thick] (q5)++(-1cm,0.05cm) -- ++(0.55cm,0cm);

    \node[below=0.2cm of ENN2] (middle) {};
    \node[right=1.7cm of middle,fill=white,draw, rounded corners,minimum height=4.5cm, minimum width=0.9cm,label={[rotate=90]center:\mbox{\large$\bm{s}' \leftarrow \mathbf{P}_\text{rand}\,\bm{s}' $}}] (UE shuf) {};

    \node[right=0.6cm of UE shuf, fill=white, draw, rounded corners, fill=gray!15, minimum height=4.5cm, minimum width=0.9cm, label={[rotate=90]center:\mbox{\large DNN with softmax}}] (DQN) {};

    \node[right=4cm of ENN4] (helper) {};
    \node[above=0.48cm of helper,draw, fill=white, rounded corners,minimum height=3.5cm, minimum width=0.9cm,label={[rotate=90]center:\mbox{\large$\bm{x} \leftarrow \mathbf{P}_\text{rand}^{-1}\,\bm{x} $}}] (UE_res) {};

    \node[draw, rectangle, below=5pt of UE shuf,fill=gray!15, rounded corners] (emb) {Embedding};
    \node at (x4|-emb) (x_prb) {$n_\text{PRB}$};
    \draw[->,thick] (x_prb) -- (emb);
    \draw[->,thick] (emb) -| (DQN);

    \node[right=2.15cm of DQN,draw, fill=white, rounded corners,minimum height=4.5cm, minimum width=0.9cm,label={[rotate=90]center:\mbox{\large Masking}}] (Mask) {};

\end{tikzpicture}
    }
    \caption{Structure of the actor network, modified from~\cite{bansbach21}. 
    ENNs, UE shuffling and PRB embedding is carried out as described in~\cite{bansbach21}.
    Learnable segments are highlighted by gray shading.
    Probabilities of taking actions~$x_k$ are returned, where~$x_{k+1}$ denotes the action of leaving the resource empty.
    The probability matrix of past action probabilities~$\bm{X}=[\bm{x}_1\ldots\bm{x}_M]$ is given as input.
    }
    \label{figure:agent}
\end{figure}

In~\cite{bansbach21}, we introduced encoder neural networks (ENNs) for effective state space compression.
The state~$\bm{s}_k$ of every UE~$k$, containing the size and ages of packets residing in a UE's buffer as well as the UE's CQI~$\mathfrak{c}_k$ and QI~$\mathfrak{q}_k$, is compressed to a vector~$\bm{s}_k' \in \mathbb{R}^3$ using ENNs.
To remedy the issue of learning a bias towards a specific UE, we randomly shuffle the order of UEs by~$\bm{s} \leftarrow \bm{P}_\text{rand}\bm{s}$, where~$\bm{P}_\text{rand}$ is a random permutation matrix.
Figure \ref{figure:agent} shows the setup of the actor network.
The compressed and randomly permuted UE information~$\bm{s}_k,\, k \in \mathcal{K}$,~$\bm{X}$ as well as~$n_\text{PRB} \in \mathcal{N}$, denoting the current PRB to allocate, are fed to a DNN with a softmax-function at its output layer.
Using~$\bm{P}_\text{rand}^{-1}$, the order of UEs is restored and the masking is applied. 
To ensure exploration during training, parameter space noise~\cite{parameternoise} is applied to the weights and biases of the output layer.
To avoid overfitting of the DNN, we use dropout layers.  \\
We furthermore use age capping for handling packets that exceeded their PDB~\cite{bansbach21}.
The critic's architecture is similar in terms of input information processing, however, the DNN's output layer only has a single neuron, since solely the expected return for executing actions~$\bm{X}$ given state~${S_t =\{\bm{s}_1,\ldots\bm{s}_K,n_\text{PRB} \}}$ needs to be estimated.

\section{Results}
\subsection{Training, Evaluation and Test Setup}
We employ the environment described in Sec. II.
For a small environment (SE) with $K=20$ UEs, $N=10$ PRBs and $W_\text{cap}=-80\,000$, we show the success of masking.
For a large environment (LE) with $K=32$ UEs, $N=25$ PRBs and $W_\text{cap}=-150\,000$, we show the necessity of dropout layers.
All setups have $L=8$ buffer slots per UE and~$M=2$.  
The parameters of the embedding and ENNs are given by~\cite[Tab.~I]{bansbach21}.
Depending on SE and LE, the DNN architecture differs according to Tab.~\ref{tab:hyperparameter}.  
The different agents we train and test are summarized in Tab.~\ref{tab:agent_overview}.
We benchmark against the \emph{NOMA uplink proportional fair channel aware} (NPFCA) agent, provided by ~\cite{NokiaWirelesssuite2021}.
Given the infinite set of environment realizations $\mathfrak{S}$, we choose $|\mathfrak{S}_\text{eval}|=4$ realizations for the evaluation and $|\mathfrak{S}_\text{test}|=100$ for the test set, with $\mathfrak{S}_\text{eval} \cap \mathfrak{S}_\text{test} =\varnothing$. 
Furthermore, $\mathfrak{S}_\text{train} \subset \mathfrak{S} \setminus (\mathfrak{S}_\text{eval} \cup \mathfrak{S}_\text{test})$ denotes the training set, where the training realizations are sampled from.
During training, an environment is stopped after $t_\text{stop}$, either after $T_\text{max}=600$ time steps are executed or by early termination, i.e., $t_\text{stop}\leq600$ time steps.
After five training episodes, the agent is evaluated.
To save computing time for LE, agents are only evaluated if they achieve a training reward better than the mean reward of the benchmark agent, which is $W_\text{eval} = -1552$.
The evaluation and tests are limited to $65\,536$ allocation steps, resulting in $T_\text{max,test}=6553$ time steps for SE and $T_\text{max,test}=2621$ time steps for LE.

\begin{table}
    \vspace{-0.3cm}
    \centering 
    \caption{Parameters of the DNNs for the SE and LE}
    \begin{tabular}{cccccc}
        \toprule 
         & input & hidden  & output & hidden & activation    \\
         & width & width   & width  & layers & functions  \\
        \midrule
        SE & $105$ & $603$ & $21$ & $3$ & ReLU    \\ 
        LE & $165$ & $963$ & $33$ & $3$ & ReLU  \\ 
        \bottomrule 
    \end{tabular}
    \label{tab:hyperparameter}
    \vspace{-0.3cm}
\end{table}

\begin{table}
    \vspace{-0.5cm}
    \centering
    \caption{Overview of the agents}
    \begin{tabular}{cccccc}
        \toprule 
            & Environment & $K$ & $N$ &Masking & Dropout  \\
        \midrule 
        S-def & SE & 20 & 10 & --- & ---\\
        S-mask & SE & 20 & 10 & $\surd$ & --- \\
        L-mask & LE & 32 & 25 & $\surd$ & --- \\
        L-drop & LE & 32 & 25 & $\surd$ & $\surd$ \\
        \bottomrule \\ 
    \end{tabular}
    \label{tab:agent_overview}
\end{table}

\subsection{Training and Evaluation Results}
\begin{figure}
    \vspace{-0.4cm}
    \centering 
    \resizebox{0.52\textwidth}{!}{
        \input{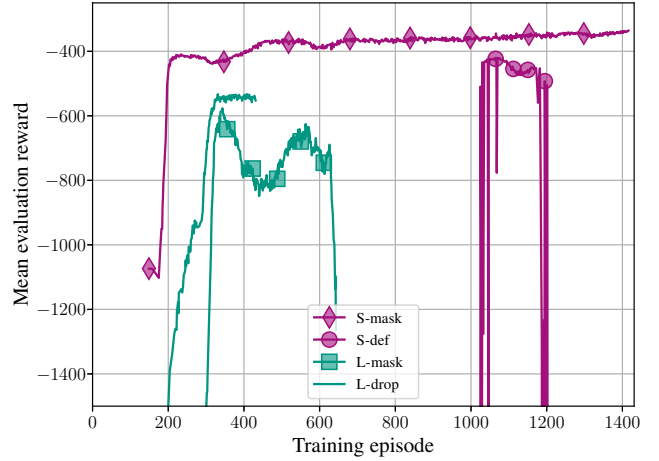}
    }   
    \caption{Mean evaluation reward of the agents introduced in Tab.~\ref{tab:agent_overview}.}
    \label{fig:training}
    \vspace{-0.4cm}
\end{figure}

\begin{figure}
    \vspace{-0.0cm}
    \centering 
    \resizebox{0.52\textwidth}{!}{
        \begin{adjustbox}{clip,trim=0cm 0cm 0cm 1.3cm}
            \input{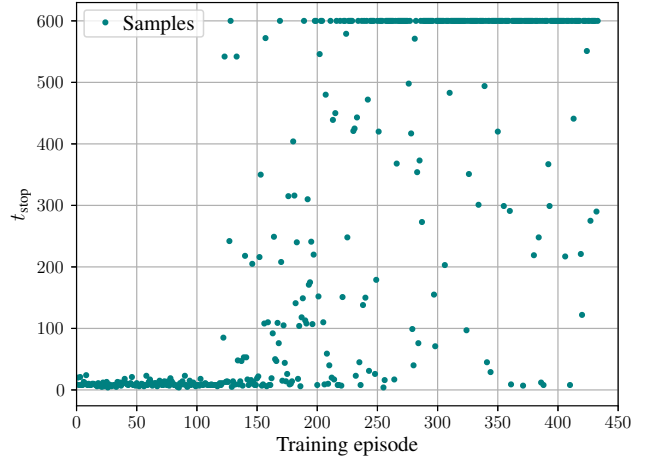}
        \end{adjustbox}
        
    }
    \caption{Maximum environment runtime $t_\text{stop}$ of the training environments of the L-drop agent when early termination is applied.
    }
    \label{fig:termination}
    \vspace{-0.3cm}
\end{figure}

In Fig.~\ref{fig:training}, the training performance of the different agents is shown.
If an agent is evaluated, its mean evaluation reward over $\mathfrak{S}_\text{eval}$ is plotted as a function of the training episode. 
The S-def agent has only a small window where its training performance is good enough to get evaluated.
With increasing training episodes, the S-mask agent is steadily improving.
We conclude that masking stabilizes and improves training significantly.
However, for the LE, the L-mask agent returns volatile evaluation rewards and its performance collapses after approximately 650 training episodes.
We assume that this is attributed to overfitting of the DNN.
Adding dropout layers with an outage probability of $p=0.2$ stabilizes training, shown by the L-drop agent. 
Due to sufficient evaluation performance and high computation effort, training of the L-drop agent was stopped after 440 training episodes. 
Figure~\ref{fig:termination} shows the impact of early termination on the training of the L-drop agent. 
With increasing training, the runtime $t_\text{stop}$ of the environment realizations increases as well.
Especially in the beginning of training, early termination significantly speeds up training.
The L-drop agent gets a performance boost at approximately 200 episodes, which, compared to Fig.~\ref{fig:training}, is the start of evaluation of the L-drop agent.

\subsection{Test Results}
\begin{figure}
    \centering 
    \vspace{-0.5cm}
    \resizebox{0.54\textwidth}{!}{
        \input{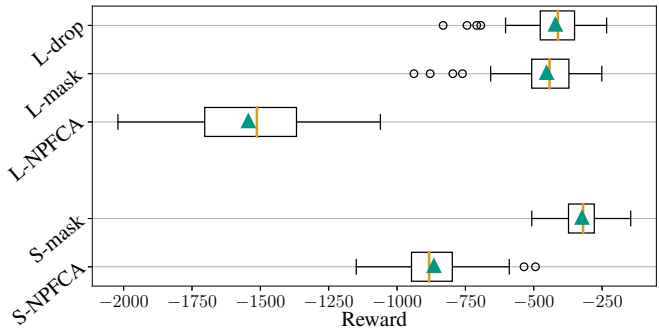}
    }
    \caption{Test rewards of the agents introduced in Tab.~\ref{tab:agent_overview} and the benchmark agents.
    The lower two agents are specialized to the SE, the upper three agents to the LE.}
    \label{fig:boxplot}
    \vspace{-0.3cm}
\end{figure}

The improvement of our proposed agents compared to the benchmark agent is shown in Fig.~\ref{fig:boxplot}.
The agents are tested using the same 100 environment realizations and we plot the obtained rewards.
Green triangles indicate the mean evaluation reward, vertical orange lines inside the box, limited by the lower and upper quartile, depict the median reward over all environments.
Open circles are outliers. 
Our agents outperform the benchmark agent by getting merely 37\% and 27\% of the benchmarks penalty, for SE and LE respectively.

\subsection{Discussion}
In \cite{bansbach21}, we have shown that incorporating ENNs compressing BSI as well as age capping enable the design of an agent that outperforms benchmark agents without BSI.
For a detailed discussion of the benefits of ENNs and age capping, we refer the interested reader to~\cite{bansbach21}.
Due to the large action space of the MC-NOMA system, the Q-learning approach from \cite{bansbach21} is not feasible anymore.
By changing the DRL technique of \cite{bansbach21} to an actor-critic approach and adding traffic-based masking, we can extend our previous work of \cite{bansbach21} and design an agent for resource allocation in an uplink MC-NOMA system using BSI. 
The proposed scheme assumes to train and test on a fixed number of UEs.
The generalization of the agent to a variable number of UEs, e.g., training for $K$ UEs and testing for $K_\text{test}<K$, is ongoing.

\section{conclusion}
In this work, we have proposed a centralized DRL agent for an uplink MC-NOMA resource allocation problem using BSI.
We proposed to use a DDPG-based approach with a stochastic policy and combine it with the methods introduced in \cite{bansbach21}.
To enable the decision for multiple actions per PRB, we proposed feedback of the probability matrix of past action to the actor.
To speed up training, we introduced early termination, which interrupts training in ill-conditioned situations.  
Furthermore, we have shown that a traffic-based masking of actions as well as dropout layers stabilize training of the agent. 
For $K=20$ and $K=32$ UEs, we significantly outperform the benchmark agent.
Thus, we enabled the use of BSI for an uplink MC-NOMA resource allocation problem, which improved the performance of the scheduler.

\bibliographystyle{IEEEtran}
\bibliography{chapters/res_alloc.bib}

\end{document}